\documentclass[aps,pre,reprint,floatfix]{revtex4-1}

\usepackage{graphicx}
\usepackage{amsmath}
\usepackage{dcolumn} 
\usepackage{multirow}

\begin{document}

\title{Predicting the structure of fluids with piecewise constant interactions: Comparing the accuracy of five efficient integral equation theories}

\author{Kyle B. Hollingshead}
\affiliation{McKetta Department of Chemical Engineering, University of Texas at Austin, Austin, Texas 78712, USA}

\author{Thomas M. Truskett}
\email[]{truskett@che.utexas.edu}
\affiliation{McKetta Department of Chemical Engineering, University of Texas at Austin, Austin, Texas 78712, USA}

\date{\today}

\begin{abstract}
We use molecular dynamics simulations to test integral equation theory predictions for the structure of fluids of spherical particles with eight different piecewise-constant pair interaction forms comprising a hard core and a combination of two shoulders and/or wells. Since model pair potentials like these are of interest for discretized or coarse-grained representations of effective interactions in complex fluids (e.g., for computationally intensive inverse optimization problems), we focus here on assessing how accurately their properties can be predicted by analytical or simple numerical closures including Percus-Yevick, hypernetted chain, reference hypernetted chain, first-order mean spherical approximation, and a modified first-order mean spherical approximation. To make quantitative comparisons between the predicted and simulated radial distribution functions, we introduce a cumulative structural error metric. For equilibrium fluid state points of these models, we find that the reference hypernetted chain closure is the most accurate of the tested approximations as characterized by this metric or related thermodynamic quantities.
\end{abstract}

\maketitle

\section{Introduction}

A common challenge in materials science is the ``inverse design problem''~\cite{torquato:invstatmech, jain:invstatmechrev}, wherein one seeks to use theoretical models to discover the microscopic characteristics (e.g., the effective pair interactions) of a new system which, if fabricated or synthesized, would yield a targeted material property. Recent applications include designing materials that self-assemble into specific crystalline lattices~\cite{jain:invstatmech, marcotte:invstatmech2, cohn:invstatmech, edlund:invstatmech}, fluids that display optimized structural correlations and related transport properties~\cite{stillinger:iso-g,carmer:tracer,goel:tuneddensity}, or solids that exhibit specific optical characteristics~\cite{andkjaer:photonics}. Inverse design problems are commonly addressed by stochastic optimization strategies like simulated annealing. Such approaches have the advantage of being general and easy to apply, and they can also be effective as long as material properties required for evaluating the objective function can be accurately and efficiently computed for large numbers of trial interactions during the optimization. This requirement typically means that ``exact'' yet computationally intensive methods for property determination (e.g., molecular simulations) are impractical for use within such calculations. Approximate theories with analytical or simple numerical solutions are attractive alternatives to molecular simulation in these contexts, provided that they can make sufficiently accurate predictions for a wide range of microscopic interaction types.

For bulk fluids, a key aim for property prediction is to discover the one-to-one link~\cite{henderson:uniqueg} between \(g(r)\), the radial distribution function (RDF) of a system at a given set of conditions, and \(\varphi(r)\), the interparticle pair potential. Knowledge of these functions of interparticle separation~$r$ allows for the direct calculation of the static structure factor, the energy, the pressure, and the isothermal compressibility~\cite{thysimpliq}. Estimations of other properties can be directly obtained from knowledge of the RDF as well. One example is the two-body excess entropy, which is often a good approximation of the total excess entropy~\cite{nettleton:s2dominant} for simple liquids. Another is the information-theoretic estimate for the probability $p_n (\Omega)$ of observing $n$ particle centers in a molecular-scale subvolume $\Omega$, a quantity which characterizes the fluid's density fluctuations~\cite{hummer:insertprob}. Excess entropy, its two-body approximation, and $p_0$ have been shown to correlate with various dynamic properties of equilibrium fluids, e.g. diffusivity or viscosity~\cite{rosenfeld:sexscaling2,dzugutov:sexscaling,rosenfeld:sexscaling1,chakravarty:waterlikesexscaling,abramson:h2osexscaling,krekelberg:s2,abramson:n2sexscaling,mittal:p0scaling,abramson:co2sexscaling,dyre:isomorphs,goel:confinedstructure,krekelberg:rosenfeld,pond:s2gauss,errington:dumbbellsexscaling,chakravarty:ionmeltsexscaling,pond:rosenfeldbrownian,carmer:tracer,bollinger:sexscaling}.
Mode-coupling theory also predicts that dynamic phenomena can be directly estimated from knowledge of the static structure factor~\cite{reichman:mctreview}. 

With these considerations in mind, herein we use molecular simulations to test the accuracy of RDF predictions for five approximate integral-equation theory closures: Percus-Yevick, hypernetted chain and reference hypernetted chain~\cite{thysimpliq}, first-order mean spherical approximation (FMSA)~\cite{tang:fmsa2}, and a modified exponential version of FMSA~\cite{hlushak:sexpfmsasquareshoulder}. 
Other more resource-intensive theories, like the Rogers-Young and hybrid mean-spherical approximations~\cite{rogers:ry, zerah:hmsa}, self-consistent Ornstein-Zernike approaches~\cite{thysimpliq}, and thermodynamic perturbation theories~\cite{barkerhenderson:swtpt,zhou:3rdordertpt,zhou:5thordertpt} are not considered here. 
We apply the simpler five theories listed above to a diverse suite of eight pair potentials previously introduced by Santos et al~\cite{santos:2steps}, each composed of a hard core at $r=\sigma$ plus two piece-wise constant sections at larger $r$ (i.e. wells or shoulders), that qualitatively mimic some of the features observed in the effective interactions of complex fluid systems.  For each interaction, we investigate four thermodynamic state points with various combinations of low and high density and low and high temperature, and we compare the theoretical predictions for the RDF, the energy, and the two-body excess entropy to data from event-driven molecular dynamics simulations. 
To facilitate the RDF comparisons we introduce a ``cumulative squared error'' metric, which provides a quantitative characterization of the overall quality of each theoretical prediction. We also assess the accuracy of predictions for the potential energy and the two-body excess entropy.

\section{Methods}
\subsection{Integral Equation Theory}
Integral equation theories for uniform, isotropic fluids typically involve solving a system of two equations: the Ornstein-Zernike relation,
\begin{equation}    
\label{eq:oz}
h(r)=c(r)+\rho\int c\left(|\mathbf{r'}-\mathbf{r}|\right)h(r')d\mathbf{r'},
\end{equation}
which defines the direct correlation function \(c(r)\) in terms of the number density $\rho$ and the total correlation function \(h(r) = g(r)-1 \), and a closure, e.g.,
\begin{equation}
\label{eq:closure}
h(r)+1=\exp\left[-\beta\varphi(r)+h(r)-c(r)+B(r)\right],
\end{equation}
which introduces the link to the pair potential \(\varphi(r)\), where \(\beta = (k_\text{B}T)^{-1}\), \(T\) is temperature, \(k_\text{B}\) is Boltzmann's constant, and \(B(r)\) is the so-called bridge function. 

Two common approximations for \(B(r)\) are the Percus-Yevick (PY) closure,
\begin{equation}
B_{\text{PY}}(r) = \ln\left[h(r)-c(r)+1\right]-h(r)+c(r),
\end{equation}
and the hypernetted chain (HNC) closure,
\begin{equation}
B_{\text{HNC}}(r) = 0.
\end{equation}

Another is the so-called reference hypernetted chain approximation (RHNC), which assumes that the bridge function can be accurately approximated by that of a reference fluid, typically one of hard spheres at the same density:
\begin{equation}
B_{\text{RHNC}}(r) = B_\text{HS}(r) .
\end{equation}
The hard-sphere fluid's bridge function \(B_\text{HS}(r)\) has been calculated through careful molecular simulations, and multiple parameterizations for its density dependence exist~\cite{malijevsky:bridgefunc,verletweis:hsrdf,grunkehenderson:hsdcf}. 
For this work, we employ the analytical parameterization proposed by Malijevsk\'{y} and Lab\'{\i}k~\cite{malijevsky:bridgefunc} for the RHNC closure.


With \(B(r)\) specified by these closures, we solve the coupled equations (\ref{eq:oz}) and (\ref{eq:closure}) using a rapidly-converging combination of Newton-Raphson and Picard root-finding methods developed by Lab\'{\i}k et al.~\cite{labik:fastthy}. 

An alternative strategy is to replace the closure of Eq.~\ref{eq:closure} with separate expressions. For example, the mean spherical approximation (MSA) assumes the following relations hold,
\begin{equation}
\begin{aligned}
g_{\text{MSA}}(r) &= 0 & \quad r < \sigma, \\
c_{\text{MSA}}(r) &= 0 & \quad r \geq \sigma. \\
\end{aligned}
\end{equation}
By further assuming first-order expansions in the characteristic dimensionless energy of the potential \(\beta\varepsilon\) for both \(g(r)\) and \(c(r)\)--e.g., \(g_{\text{FMSA}}(r) = g_\text{HS}(r) + \beta\varepsilon g_1(r)\), where \(g_\text{HS}(r)\) is the pair correlation function for a hard sphere system at the same density--Tang and Lu closed the equations analytically for several common pair interactions, including square wells~\cite{tang:fmsa2}. We refer to this solution as the first-order mean spherical approximation (FMSA). In principle, FMSA can be applied to potentials with square shoulders as well. But for strong interactions, FMSA is known to incorrectly predict RDFs with negative values for some interparticle separations~\cite{hlushak:sexpfmsasquareshoulder}. 
To resolve this, Hlushak et al. modified the FMSA to make it equally applicable to wells and shoulders by rearranging the terms in the series expansion, so that \(g_{\text{EFMSA}}(r)=g_\text{HS}(r)\exp[-\beta\varepsilon g_1(r)]\)~\cite{hlushak:sexpfmsasquareshoulder}. In this work, we refer to this analytical solution as the exponential first-order mean spherical approximation (EFMSA).

\subsection{Suite of Two-Step Potentials}

Motivated by Santos et al.~\cite{santos:2steps}, we examine predictions for fluids from a set of pair interactions comprising a hard core and two piecewise-constant steps, 
\begin{equation}
\label{eq-potential}
\varphi(r) = \left\{ \begin{array}{ll}
\infty & \quad r < \sigma, \\
\varepsilon_1 & \quad \sigma \leq r < \lambda_1, \\ 
\varepsilon_2 & \quad \lambda_1 \leq r < \lambda_2, \\ 
0 & ~~r \ge \lambda_2, \\
\end{array} \right.
\end{equation}
where \(\varepsilon_1\) and \(\varepsilon_2\) are the energies of the first and second steps, respectively, and \(\lambda_1\) and \(\lambda_2\) are the outer edges of the first and second steps, respectively. 

\begin{figure}
\centering
\includegraphics[width=8cm]{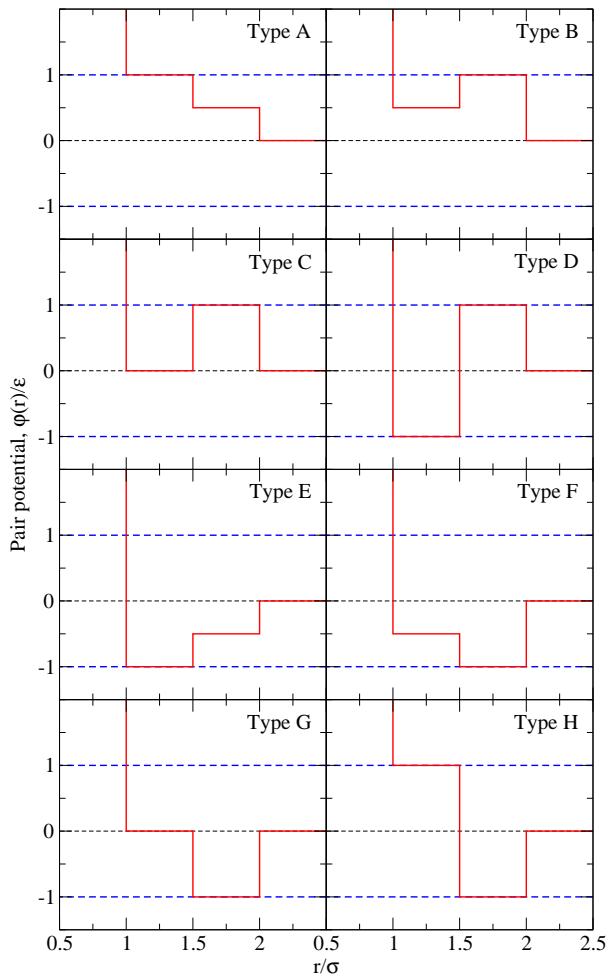}
\caption{\label{fig:interactions} The suite of eight pair interactions considered in this study, inspired by Santos et al.~\cite{santos:2steps}, is topologically exhaustive (e.g., there are no other qualitative arrangements of two constant pairwise pieces that are not more appropriately labeled single wells or shoulders).} 
\end{figure}

Furthermore, as in Santos et al., we restrict the values of \(\varepsilon_i\) to the set \(\{-\varepsilon, -\varepsilon/2, 0, \varepsilon/2, \varepsilon\}\), where \(\varepsilon\) is a characteristic energy scale. 
Cases where \(\varepsilon_1=\varepsilon_2\) or \(\varepsilon_2 = 0\) reduce to either single square wells or shoulders, or hard spheres, which have all been studied extensively elsewhere (see, e.g., refs. 1-41 in~\cite{yuste:rfass}) and are not considered here. 
We also exclude cases where \(\max\{|\varepsilon_1|, |\varepsilon_2|\} = \varepsilon/2\). Of the cases where \(\varepsilon_1\) and \(\varepsilon_2\) have opposite sign, we consider only combinations where \(\varepsilon_2=-\varepsilon_1=\pm\varepsilon\). 
We choose \(\lambda_1=1.5\sigma\) and \(\lambda_2=2\sigma\) in order to provide challenging perturbations to the bare hard sphere system that are still amenable to molecular simulation and theoretical treatment. After imposing these restrictions, the remaining eight pairwise interactions shown in Fig.~\ref{fig:interactions}, which we refer to as ``Type A--H,'' form our test suite.

To explore how the accuracy of the various theories varies with density and temperature, we investigate each interaction at the four state points comprising combinations of packing fraction \(\eta=\rho\pi\sigma^3/6=0.15\) or \(0.45\) and dimensionless temperature \(T^*=k_\text{B}T/\varepsilon=0.67\) or \(2.0\).

\subsection{Molecular Simulations}

We compare the theoretical predictions for the RDF, the energy, and the two-body excess entropy to the results of event-driven molecular dynamics simulations performed with the DynamO simulation engine~\cite{bannerman:dynamo}. Periodic boundary conditions were used, and the simulated systems were sized such that adequate RDF statistics could be collected for separations up to at least \(r=10\sigma\). In practice, this required \(N=4000\) particles when \(\eta=0.15\), and \(N=8788\) particles when \(\eta=0.45\). The ``bins'' for particle counts were \(0.005\sigma\) wide. Temperatures were set and maintained using an Andersen thermostat~\cite{andersen:thermostat}.

Each simulation was initialized as an FCC lattice of the desired density at a high temperature, with randomly assigned particle velocities. After equilibrating for ten million events, the simulations were cooled to the desired temperature and re-equilibrated for a further ten million events. Then, the thermostat was removed, and the RDF was measured over the final five million events.

\subsection{Quantifying Error in Predictions}

To compare the various RDF theoretical predictions to simulations at a given state point, we define a metric we call the cumulative squared error, \(\text{CSE}(r)\):

\begin{equation}
\label{eq:CSE}
\text{CSE}(r)=\frac{\int_{\sigma}^r \left[h_\text{sim}(r')-h_\text{thy}(r')\right]^2r'^2\,dr'}
{\int_{\sigma}^\infty h_\text{sim}^2(r')r'^2\,dr'}.
\end{equation}

The integrand in the numerator characterize the squared deviation in the total correlation function between the prediction of a given theory $h_\text{thy} (r)$ and the result of the `exact' simulation $h_\text{sim} (r)$; the power of two eliminates any possible cancellation of error, e.g. for cases where a theory both underpredicts and overpredicts the value of \(h(r)\) at different values of \(r\). The denominator accumulates the total squared correlations in the simulated system, and thus normalizes the overall function to facilitate comparison between systems with different degrees of correlation (e.g., between low-density and high-density systems).

As \(r\) approaches infinity, all \(h(r)\) curves converge to zero and the \(\text{CSE}\) converges to a finite value, \(\text{CSE}_\infty\):
\begin{equation}
\text{CSE}_\infty = \lim_{r \to \infty}{\text{CSE}(r)},
\end{equation}
which is a measure of the summed squared correlations as a fraction of the total squared correlations in the system; thus, a larger value of \(\text{CSE}_\infty\) indicates that a theoretical prediction deviates more significantly from the ``exact'' simulation results. By construction, \(\text{CSE}_\infty\) has a defined minimum of \(0\) and, while it does not have a rigorous maximum, its value is typically less than \(1\) except in cases where the theoretical predictions are qualitatively very poor.

We also calculate the potential energy per particle \(U/\varepsilon\),
\begin{equation}
\label{eq:potenergy}
\frac{U}{\varepsilon} = \frac{\rho}{2}\int_0^\infty \frac{\varphi(r)}{\varepsilon} g(r) \, d{\bf r},
\end{equation}
and the two-body contribution to excess entropy \(s^{(2)}/k_\text{B}\),
\begin{equation}
\label{eq:s2}
\frac{s^{(2)}}{k_\text{B}} = -\frac{\rho}{2}\int_0^\infty \left[g(r)\ln g(r) - g(r) + 1 \right] \, d{\bf r},
\end{equation}
from simulations and theoretical predictions. Both quantities can also be directly computed from \(g(r)\) and thus, the normalized absolute deviation of the predicted versus simulated values can be used as an indication of the success of theoretical predictions. However, note that different RDFs can, in principle, give rise to the same value of \(U/\varepsilon\) or \(s^{(2)}/k_\text{B}\). Moreover, \(U/\varepsilon\) only depends on correlations within the range of the pair interaction. As a result, we argue here that since the RDF is weighted differently for each thermodynamic quantity, the CSE metric we introduce--which tests the overall similarity between predicted and simulated RDFs--represents a more sensitive measure for the overall predictive quality of particular theory.

\begin{figure}
\centering
\includegraphics[width=8cm]{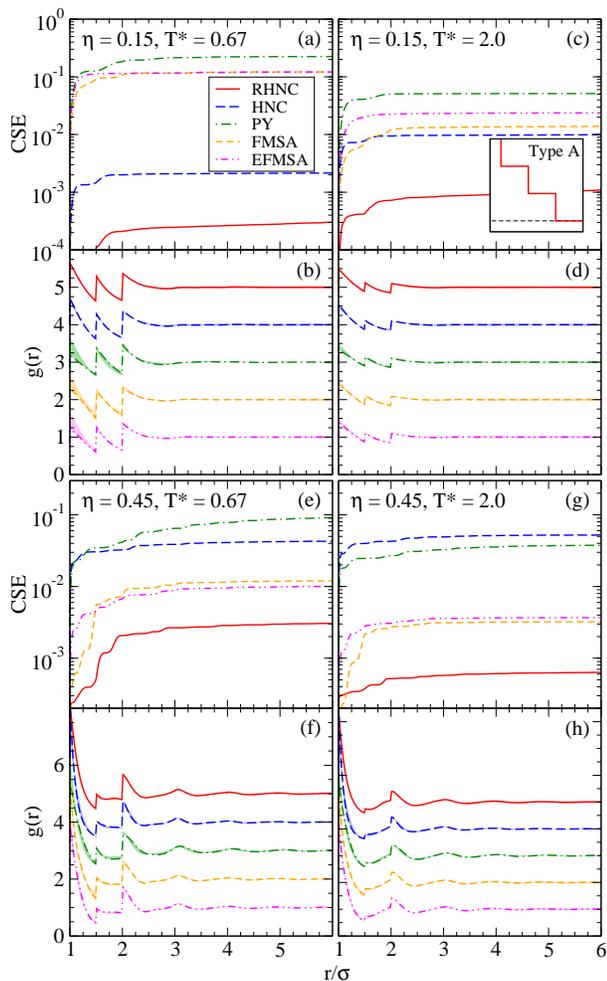}
\caption{\label{fig:adata} 
Radial distribution functions \(g(r)=h(r)+1\) and the associated cumulative squared errors (CSE, see Eq. (\ref{eq:CSE})) predicted by the reference hypernetted chain (RHNC),  hypernetted chain (HNC), and Percus-Yevick (PY) Ornstein-Zernike closures~\cite{thysimpliq,malijevsky:bridgefunc}; the first-order mean spherical approximation solution (FMSA)~\cite{tang:fmsa2}; and the simple exponential first-order mean spherical approximation (EFMSA)~\cite{hlushak:sexpfmsasquareshoulder}, for the ``type A'' pair interaction. Shaded regions adjacent to each \(g(r)\) indicate the difference between the theory and simulation results.}
\end{figure}

\section{Results and Discussion}
Structural predictions for the Type A pair interaction are compared to simulation results in Fig.~\ref{fig:adata}, along with the corresponding cumulative squared errors as calculated via Eq.~(\ref{eq:CSE}). 
For this interaction, the analytic solutions (FMSA and EFMSA) perform better at higher rather than at lower equilibrium fluid densities. As density increases, the effect of the excluded volume captured by the well-modeled hard-sphere RDF, \(g_\text{HS}(r)\), overwhelm the energetic perturbations from the repulsive steps and dominate the resulting structure. Of the tested integral-equation theories with simple numerical closures, the PY closure tends to perform least well near contact, and for interaction Type A, the RHNC offers the best predictions at all four state points investigated. Analogous figures for each of the other interactions are presented for the interested reader in Appendix~\ref{app:plots}.

It is tempting to conclude from a visual comparison of theoretical and simulated radial distribution functions that all of the theories perform similarly well, especially at the higher temperature (Figs.~\ref{fig:adata}d and \ref{fig:adata}h). However, the resulting CSEs differ {\em by nearly two orders of magnitude} from most to least accurate (Figs.~\ref{fig:adata}c and \ref{fig:adata}g), which underscores the utility and sensitivity of the CSE metric. As discussed below, these differences in the CSE become important when computing other quantities that depend on the RDF, especially when one considers that each thermodynamic quantity weights the RDF in a different way.

\begin{table}
\caption{\label{table:cse}
Total cumulative squared errors (\(\text{CSE}_\infty\)) for all theoretical approaches, thermodynamic state points, and interactions considered. ``R,'' ``H,'' and ``P'' are the RHNC~\cite{thysimpliq,malijevsky:bridgefunc}, HNC~\cite{thysimpliq}, and PY~\cite{thysimpliq} closures to the Ornstein-Zernike relation, respectively. ``F'' is the FMSA~\cite{tang:fmsa2}, and ``E'' is the EFMSA~\cite{hlushak:sexpfmsasquareshoulder}. Italics indicate the lowest value of \(\text{CSE}_\infty\) (and hence the  theory with the most accurate structural prediction) at each combination of state point and interaction type.}
\begin{ruledtabular}
\begin{tabular}{ccccccccccc}
& & \multicolumn{4}{c}{Type A} & & \multicolumn{4}{c}{Type B} \\ 
\cline{3-6} \cline{8-11}
\(T^*\) & & \multicolumn{2}{c}{0.67} & \multicolumn{2}{c}{2.00} & & 
\multicolumn{2}{c}{0.67} & \multicolumn{2}{c}{2.00} \\ 
\(\eta\) & & 0.15 & 0.45 & 0.15 & 0.45 & & 
0.15 & 0.45 & 0.15 & 0.45 \\ 
\hline 
R & & \emph{0.000}
& \emph{0.003}
& \emph{0.002}
& \emph{0.001} 
& 
& \emph{0.001} 
& \emph{0.011} 
& \emph{0.000} 
& \emph{0.002} 
\\ 
H & & 0.002 
& 0.043 
& 0.010 
& 0.052 
& 
& 0.001
& 0.037 
& 0.001 
& 0.057 
\\ 
P & & 0.225 
& 0.095 
& 0.052 
& 0.038 
& 
& 0.013 
& 0.021 
& 0.005 
& 0.007 
\\ 
F & & 0.122 
& 0.012 
& 0.014 
& 0.003 
& 
& 0.092 
& 0.053 
& 0.010 
& 0.006 
\\ 
E & & 0.121 
& 0.010 
& 0.024 
& 0.004 
& 
& 0.082 
& 0.098 
& 0.008 
& 0.011 
\\ 
\hline 
\multirow{11}{*}{} \\ 

& & \multicolumn{4}{c}{Type C} & & \multicolumn{4}{c}{Type D} \\ 
\cline{3-6} \cline{8-11}
\(T^*\) & & \multicolumn{2}{c}{0.67} & \multicolumn{2}{c}{2.00} & & 
\multicolumn{2}{c}{0.67} & \multicolumn{2}{c}{2.00} \\ 
\(\eta\) & & 0.15 & 0.45 & 0.15 & 0.45 & & 
0.15 & 0.45 & 0.15 & 0.45 \\ 
\hline 
R & & \emph{0.002}
& \emph{0.010} 
& \emph{0.000}
& \emph{0.003} 
& 
& \emph{0.079}
& 0.047 
& \emph{0.001} 
& \emph{0.004} 
\\ 
H & & 0.003 
& 0.022 
& 0.001 
& 0.059 
& 
& 0.082 
& \emph{0.031}
& 0.002 
& 0.055 
\\ 
P & & 0.005 
& 0.019 
& 0.001 
& 0.003 
& 
& 0.092 
& 0.203 
& 0.002 
& 0.020 
\\ 
F & & 0.096 
& 0.082 
& 0.008 
& 0.009 
& 
& 0.168 
& 0.276 
& 0.016 
& 0.018 
\\ 
E & & 0.184 
& 0.159 
& 0.015 
& 0.016 
& 
& 0.411 
& 0.524 
& 0.040 
& 0.032 
\\ 
\hline 
\multirow{11}{*}{} \\ 

& & \multicolumn{4}{c}{Type E} & & \multicolumn{4}{c}{Type F} \\ 
\cline{3-6} \cline{8-11}
\(T^*\) & & \multicolumn{2}{c}{0.67} & \multicolumn{2}{c}{2.00} & & 
\multicolumn{2}{c}{0.67} & \multicolumn{2}{c}{2.00} \\ 
\(\eta\) & & 0.15 & 0.45 & 0.15 & 0.45 & & 
0.15 & 0.45 & 0.15 & 0.45 \\ 
\hline 
R & & --\footnotemark[1]
& --\footnotemark[1]
& --\footnotemark[1]
& \emph{0.001} 
& 
& --\footnotemark[1]
& 0.082 
& --\footnotemark[1]
& \emph{0.004} 
\\ 
H & & --\footnotemark[1]
& --\footnotemark[1]
& --\footnotemark[1]
& 0.067 
& 
& --\footnotemark[1]
& \emph{0.027} 
& --\footnotemark[1]
& 0.049 
\\ 
P & & --\footnotemark[1]
& --\footnotemark[1]
& --\footnotemark[1]
& 0.020 
& 
& --\footnotemark[1]
& 0.255 
& --\footnotemark[1]
& 0.045 
\\ 
F & & --\footnotemark[1]
& --\footnotemark[1]
& --\footnotemark[1]
& 0.004 
& 
& --\footnotemark[1]
& 0.275 
& --\footnotemark[1]
& 0.026 
\\ 
E & & --\footnotemark[1]
& --\footnotemark[1]
& --\footnotemark[1]
& 0.002 
& 
& --\footnotemark[1]
& 0.113 
& --\footnotemark[1]
& 0.012 
\\ 
\hline 
\multirow{11}{*}{} \\ 

& & \multicolumn{4}{c}{Type G} & & \multicolumn{4}{c}{Type H} \\ 
\cline{3-6} \cline{8-11}
\(T^*\) & & \multicolumn{2}{c}{0.67} & \multicolumn{2}{c}{2.00} & & 
\multicolumn{2}{c}{0.67} & \multicolumn{2}{c}{2.00} \\ 
\(\eta\) & & 0.15 & 0.45 & 0.15 & 0.45 & & 
0.15 & 0.45 & 0.15 & 0.45 \\ 
\hline 
R & & --\footnotemark[1]
& 0.018 
& \emph{0.001} 
& \emph{0.006} 
& 
& \emph{0.052} 
& 0.043 
& \emph{0.004} 
& \emph{0.012} 
\\ 
H & & --\footnotemark[1]
& \emph{0.011} 
& 0.008 
& 0.041 
& 
& 0.063 
& \emph{0.026} 
& 0.008 
& 0.029 
\\ 
P & & --\footnotemark[1]
& 0.391 
& 0.007 
& 0.087 
& 
& 0.104 
& 0.820 
& 0.008 
& 0.204 
\\ 
F & & --\footnotemark[1]
& 0.253 
& 0.018 
& 0.038 
& 
& 0.208 
& 0.526 
& 0.032 
& 0.078 
\\ 
E & & --\footnotemark[1]
& 0.196 
& 0.020 
& 0.020 
& 
& 0.750 
& 0.485 
& 0.044 
& 0.048 
\\ 
\end{tabular}

\end{ruledtabular}
\footnotetext[1]{Simulated system is not a single-phase, uniform fluid at equilibrium.}
\end{table}

The total cumulative squared errors \(\text{CSE}_\infty\) for all interactions, state points, and theories are listed in Table \ref{table:cse}.
Six of the total thirty-two combinations of interaction type and state point considered did not produce single-phase, uniform fluids when simulated. Of the remaining twenty-six systems, the RHNC offered the most accurate structural predictions for all but four; however, at three of these four points, the \(\text{CSE}_\infty\) of the RHNC is still within ca. 65\% of the most accurate theory (HNC). All four points are at low temperature (\(T^*=0.67\)) and high packing fraction (\(\eta=0.45\)), and each of the pair interactions include attractions (types D, F, G, and H). 

\begin{figure}
\centering
\includegraphics[width=8cm]{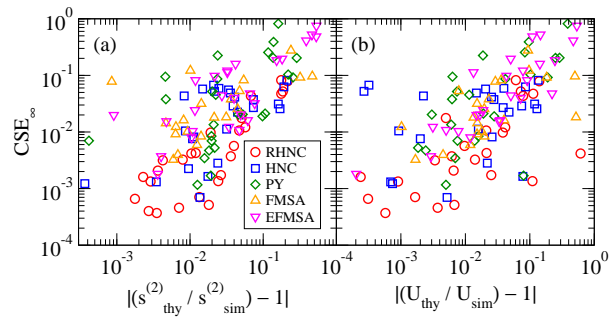}
\caption{\label{fig:corr} 
Correlations between total cumulative squared error \(\text{CSE}_\infty\) and either (a) absolute normalized two-body excess entropy error or (b) absolute normalized potential energy error for all data collected.}
\end{figure}

We also compare \(\text{CSE}_\infty\) against the absolute normalized errors for predictions of two example thermodynamic quantities, two-body excess entropy \(s^{(2)}/k_\text{B}\) and potential energy \(U/\varepsilon\), in Fig.~\ref{fig:corr}. Fig.~\ref{fig:corr}a shows that \(\text{CSE}_\infty\) is generally a good predictor of \(s^{(2)}/k_\text{B}\) accuracy, although there are a handful of instances where the fractional error in the excess entropy is low while \(\text{CSE}_\infty\) is higher.The correlation between \(\text{CSE}_\infty\) and the potential energy is weaker, but still present; this is likely due to opportunities for fortuitous cancellation of error when pair interactions contain both positive and negative contributions (e.g., types D and H), when portions of the interactions are zero (types C and G), or when significant contributions to \(\text{CSE}_\infty\) occur beyond the range of the pair interaction. Overall, however, it is clear that the accuracies of both example thermodynamic quantity predictions correlate well with the cumulative squared error. For the interested reader, the values of \(\left|\left(s^{(2)}_\text{thy}/s^{(2)}_\text{sim}\right)-1\right|\) and \(\left|\left(U_\text{thy}/U_\text{sim}\right)-1\right|\) are tabulated in Appendix~\ref{app:tables}. If other thermodynamic quantities that depend on the RDF in a different way (e.g., the pressure or the isothermal compressibility) are also of interest, then the necessity to have an independent structural metric like \(\text{CSE}_\infty\) to assess the quality of the structural predictions is even more critical.

\section{Conclusion}
In order to quantify the overall accuracy of theoretical predictions for fluid structure, we have introduced the total cumulative squared error (\(\text{CSE}_\infty\)) metric, which accumulates squared discrepancies between a theoretical prediction and a reference ``exact'' result at all separation distances along the total correlation function and avoids any possible cancellation of error. We find that this \(\text{CSE}_\infty\) metric is very sensitive and tends to forecast the overall accuracy of structure-dependent thermodynamic calculations.  As a result, it is an excellent tool for comparing accuracy between multiple theories, particularly when differences are difficult to discern by visual inspection.

We have used this metric to test the performance of five integral equation theory-based approaches for predicting equilibrium fluid structure in systems with pair interactions comprising a hard core plus two piecewise constant interactions, and we find that the reference hypernetted chain (RHNC) integral equation closure offers accurate and efficient predictions across a broad range of interactions and thermodynamic state points. This kind of analysis, i.e., considering the accuracy of various efficient theoretical methods for predicting the structure consistent with a broad range of possible interactions, will be particularly important for inverse design problems where the goal is to rather accurately predict which interaction is consistent with a targeted structure (or structurally-related property).


\begin{acknowledgments}
The authors thank Anatol Malijevsk\'y for sharing his rapidly-converging integral equation theory code. 
T.M.T. acknowledges support of the Welch Foundation (F-1696) and the National Science Foundation (CBET-1403768). 
We also acknowledge the Texas Advanced Computing Center (TACC) at The University of Texas at Austin for providing HPC resources that have contributed to the research findings reported within this paper.
\end{acknowledgments}

\appendix

\section{Extended Type B-H Structure Plots}
\label{app:plots}

\begin{figure}[H]
\centering
\includegraphics[width=8cm]{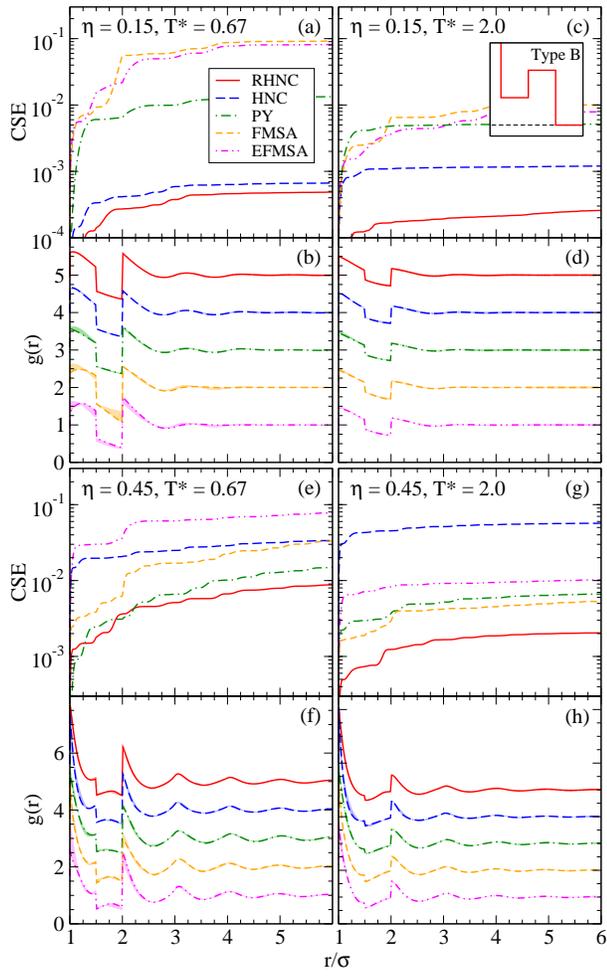}
\caption{\label{fig:bdata} 
Radial distribution functions \(g(r)=h(r)+1\) and the associated cumulative squared errors (CSE, see Eq. (\ref{eq:CSE})) predicted by the reference hypernetted chain (RHNC),  hypernetted chain (HNC), and Percus-Yevick (PY) Ornstein-Zernike closures~\cite{thysimpliq,malijevsky:bridgefunc}; the first-order mean spherical approximation solution (FMSA)~\cite{tang:fmsa2}; and the simple exponential first-order mean spherical approximation (EFMSA)~\cite{hlushak:sexpfmsasquareshoulder}, for the ``type A'' pair interaction. Shaded regions adjacent to each \(g(r)\) indicate the difference between the theory and simulation results.}
\end{figure}

\begin{figure}[H]
\includegraphics[width=8cm]{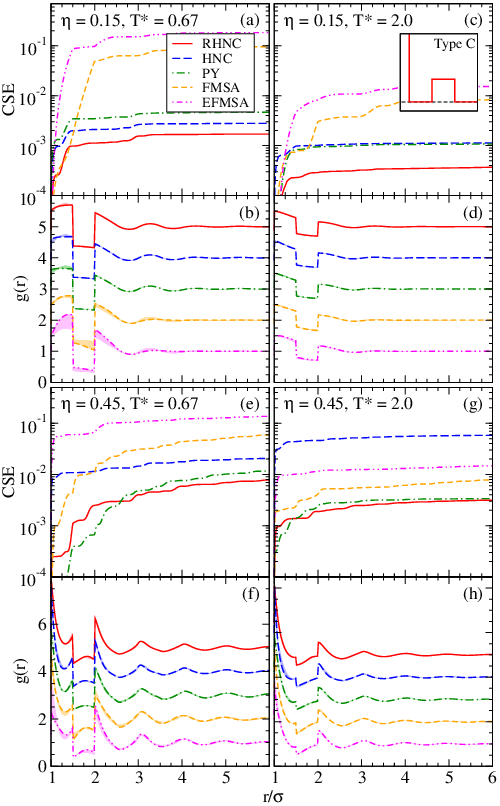}
\caption{\label{fig:cdata} Radial distribution functions and cumulative squared errors for the ``type C'' interaction. Series and labeling are as in Fig. \ref{fig:bdata}.}
\end{figure}

\begin{figure}[H]
\includegraphics[width=8cm]{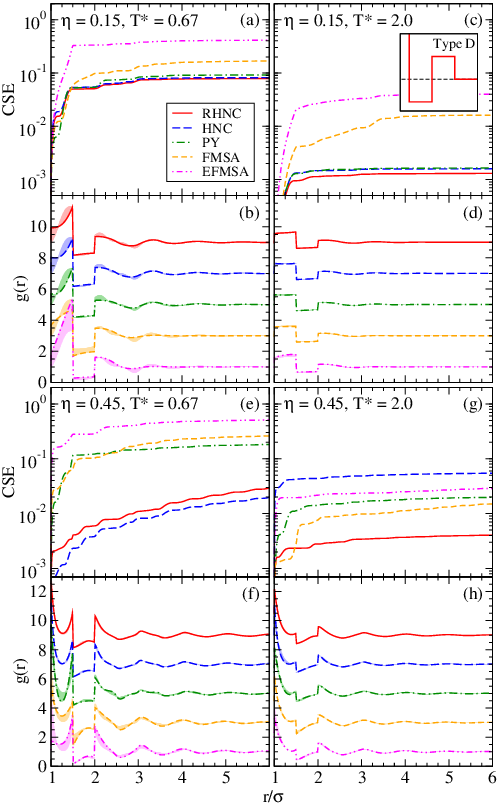}
\caption{\label{fig:ddata} Radial distribution functions and cumulative squared errors for the ``type D'' interaction. Series and labeling are as in Fig. \ref{fig:bdata}.}
\end{figure}

\begin{figure}[H]
\centering
\includegraphics[width=4.5cm]{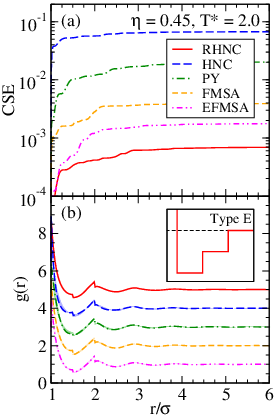}
\caption{\label{fig:edata} Radial distribution functions and cumulative squared errors for the ``type E'' interaction. Series and labeling are as in Fig. \ref{fig:bdata}.}
\end{figure}

\begin{figure}[H]
\includegraphics[width=8cm]{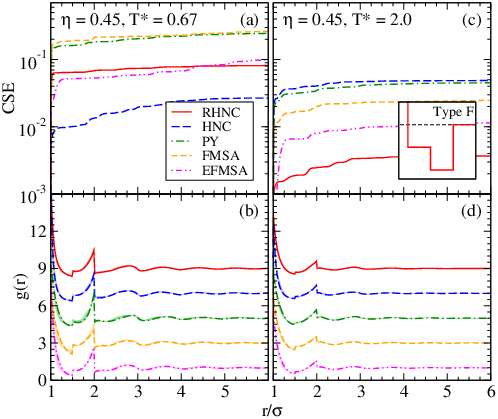}
\caption{\label{fig:fdata} Radial distribution functions and cumulative squared errors for the ``type F'' interaction. Series and labeling are as in Fig. \ref{fig:bdata}.}
\end{figure}

\begin{figure}[H]
\includegraphics[width=8cm]{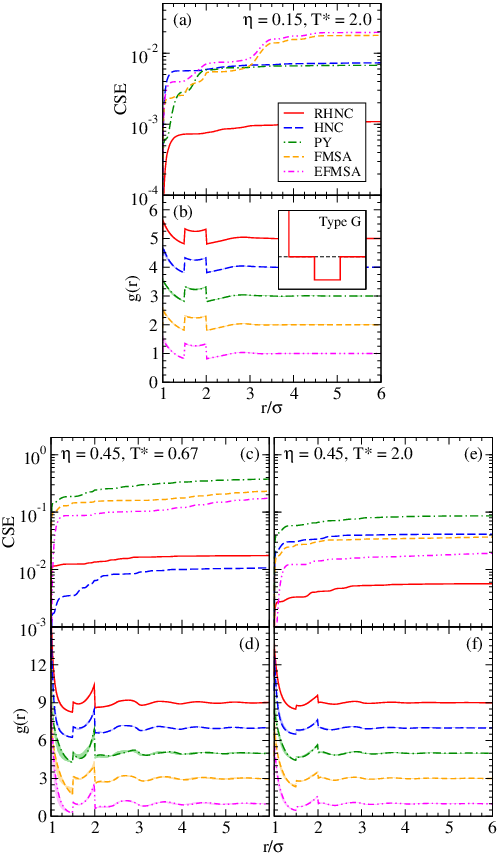}
\caption{\label{fig:gdata} Radial distribution functions and cumulative squared errors for the ``type G'' interaction. Series and labeling are as in Fig. \ref{fig:bdata}.}
\end{figure}

\begin{figure}[H]
\includegraphics[width=8cm]{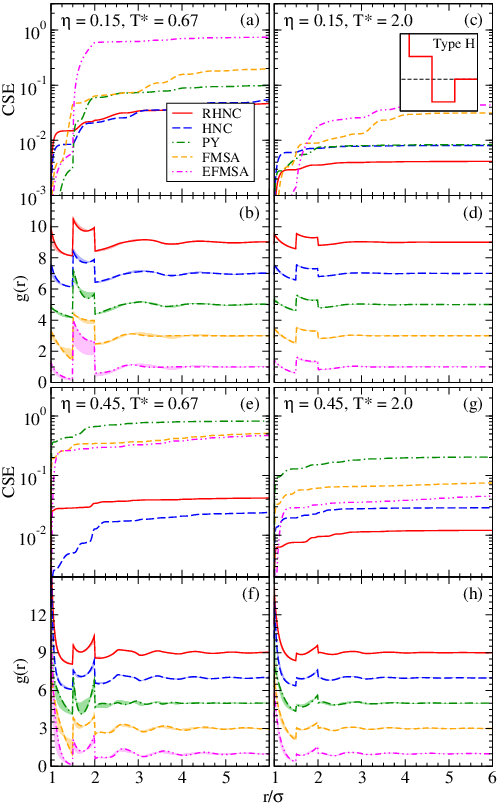}
\caption{\label{fig:hdata} Radial distribution functions and cumulative squared errors for the ``type H'' interaction. Series and labeling are as in Fig. \ref{fig:bdata}.}
\end{figure}

The predicted radial distribution functions \(g(r)\) compared against simulation results, and the resulting cumulative squared errors \(\text{CSE}(r)\), are shown for interaction types B through H in Figs. \ref{fig:bdata}--\ref{fig:hdata}, respectively.

\section{Complete Thermodynamic Error Tables}
\label{app:tables}

\begin{table}[H]
\caption{\label{table:potenergy}
Absolute normalized potential energy error, \(|(U_\text{thy}/U_\text{sim})-1|\), for all approaches, state points, and interactions considered. Labels are as in Table~\ref{table:cse}. Italics indicate the value closest to zero (e.g., a perfect prediction) at each combination of state point and interaction type.}
\begin{ruledtabular}
\begin{tabular}{ccccccccccc}
& & \multicolumn{4}{c}{Type A} & & \multicolumn{4}{c}{Type B} \\ 
\cline{3-6} \cline{8-11}
\(T^*\) & & \multicolumn{2}{c}{0.67} & \multicolumn{2}{c}{2.00} & & 
\multicolumn{2}{c}{0.67} & \multicolumn{2}{c}{2.00} \\ 
\(\eta\) & & 0.15 & 0.45 & 0.15 & 0.45 & & 
0.15 & 0.45 & 0.15 & 0.45 \\ 
\hline 
R & & \emph{0.000} 
& 0.003 
& \emph{0.000} 
& \emph{0.000} 
& 
& 0.007 
& 0.021 
& \emph{0.001} 
& 0.007 
\\ 
H & & 0.000 
& 0.002 
& 0.001 
& 0.000 
& 
& 0.005 
& 0.028 
& 0.001 
& 0.021 
\\ 
P & & 0.019 
& 0.006 
& 0.007 
& 0.006 
& 
& 0.006 
& 0.019 
& 0.002 
& 0.008 
\\ 
F & & 0.097 
& 0.001 
& 0.015 
& 0.002 
& 
& 0.182 
& 0.008 
& 0.018 
& 0.012 
\\ 
E & & 0.079 
& 0.008 
& 0.019 
& 0.003 
& 
& 0.091 
& 0.043 
& 0.012 
& 0.005 
\\ 
\hline 
\multirow{11}{*}{} \\ 

& & \multicolumn{4}{c}{Type C} & & \multicolumn{4}{c}{Type D} \\ 
\cline{3-6} \cline{8-11}
\(T^*\) & & \multicolumn{2}{c}{0.67} & \multicolumn{2}{c}{2.00} & & 
\multicolumn{2}{c}{0.67} & \multicolumn{2}{c}{2.00} \\ 
\(\eta\) & & 0.15 & 0.45 & 0.15 & 0.45 & & 
0.15 & 0.45 & 0.15 & 0.45 \\ 
\hline 
R & & 0.025 
& 0.038 
& 0.004 
& 0.014 
& 
& 0.140 
& 0.113 
& 0.110 
& 0.613 
\\ 
H & & 0.022 
& 0.048 
& 0.001 
& 0.040 
& 
& 0.135 
& 0.119 
& 0.078 
& 1.698 
\\ 
P & & 0.002 
& 0.028 
& 0.005 
& 0.007 
& 
& 0.189 
& 0.250 
& 0.080 
& 2.030 
\\ 
F & & 0.501 
& 0.036 
& 0.020 
& 0.015 
& 
& 0.181 
& 0.610 
& 0.040 
& 0.520 
\\ 
E & & 0.245 
& 0.143 
& 0.034 
& 0.035 
& 
& 0.472 
& 0.146 
& 0.072 
& 3.431 
\\ 
\hline 
\multirow{11}{*}{} \\ 

& & \multicolumn{4}{c}{Type E} & & \multicolumn{4}{c}{Type F} \\ 
\cline{3-6} \cline{8-11}
\(T^*\) & & \multicolumn{2}{c}{0.67} & \multicolumn{2}{c}{2.00} & & 
\multicolumn{2}{c}{0.67} & \multicolumn{2}{c}{2.00} \\ 
\(\eta\) & & 0.15 & 0.45 & 0.15 & 0.45 & & 
0.15 & 0.45 & 0.15 & 0.45 \\ 
\hline 
R & & --\footnotemark[1] 
& --\footnotemark[1] 
& --\footnotemark[1] 
& 0.002 
& 
& --\footnotemark[1] 
& 0.061 
& --\footnotemark[1] 
& 0.006 
\\ 
H & & --\footnotemark[1] 
& --\footnotemark[1] 
& --\footnotemark[1] 
& 0.000 
& 
& --\footnotemark[1] 
& 0.039 
& --\footnotemark[1] 
& 0.015 
\\ 
P & & --\footnotemark[1] 
& --\footnotemark[1] 
& --\footnotemark[1] 
& 0.003 
& 
& --\footnotemark[1] 
& 0.084 
& --\footnotemark[1] 
& 0.011 
\\ 
F & & --\footnotemark[1] 
& --\footnotemark[1] 
& --\footnotemark[1] 
& 0.005 
& 
& --\footnotemark[1] 
& 0.094 
& --\footnotemark[1] 
& 0.020 
\\ 
E & & --\footnotemark[1] 
& --\footnotemark[1] 
& --\footnotemark[1] 
& 0.000 
& 
& --\footnotemark[1] 
& 0.013 
& --\footnotemark[1] 
& 0.003 
\\ 
\hline 
\multirow{11}{*}{} \\ 

& & \multicolumn{4}{c}{Type G} & & \multicolumn{4}{c}{Type H} \\ 
\cline{3-6} \cline{8-11}
\(T^*\) & & \multicolumn{2}{c}{0.67} & \multicolumn{2}{c}{2.00} & & 
\multicolumn{2}{c}{0.67} & \multicolumn{2}{c}{2.00} \\ 
\(\eta\) & & 0.15 & 0.45 & 0.15 & 0.45 & & 
0.15 & 0.45 & 0.15 & 0.45 \\ 
\hline 
R & & --\footnotemark[1] 
& 0.005 
& 0.001 
& 0.010 
& 
& 0.073 
& 0.079 
& 0.021 
& 0.041 
\\ 
H & & --\footnotemark[1] 
& 0.016 
& 0.002 
& 0.025 
& 
& 0.086 
& 0.129 
& 0.026 
& 0.075 
\\ 
P & & --\footnotemark[1] 
& 0.096 
& 0.014 
& 0.042 
& 
& 0.103 
& 0.387 
& 0.016 
& 0.289 
\\ 
F & & --\footnotemark[1] 
& 0.026 
& 0.013 
& 0.023 
& 
& 0.140 
& 0.200 
& 0.015 
& 0.029 
\\ 
E & & --\footnotemark[1] 
& 0.051 
& 0.015 
& 0.017 
& 
& 0.536 
& 0.110 
& 0.051 
& 0.221 
\\ 
\end{tabular}

\end{ruledtabular}
\footnotetext[1]{Simulated system is not a single-phase, uniform fluid at equilibrium.}
\end{table}

\begin{table}[H]
\caption{\label{table:s2}
Absolute normalized 2-body excess entropy error, \(|(s^{(2)}_\text{thy}/s^{(2)}_\text{sim})-1|\), for all approaches, state points, and interactions considered. Labels are as in Table~\ref{table:cse}. Italics indicate the value closest to zero (e.g., a perfect prediction) at each combination of state point and interaction type.}
\begin{ruledtabular}
\begin{tabular}{ccccccccccc}
& & \multicolumn{4}{c}{Type A} & & \multicolumn{4}{c}{Type B} \\ 
\cline{3-6} \cline{8-11}
\(T^*\) & & \multicolumn{2}{c}{0.67} & \multicolumn{2}{c}{2.00} & & 
\multicolumn{2}{c}{0.67} & \multicolumn{2}{c}{2.00} \\ 
\(\eta\) & & 0.15 & 0.45 & 0.15 & 0.45 & & 
0.15 & 0.45 & 0.15 & 0.45 \\ 
\hline 
R & & \emph{0.003} 
& \emph{0.004} 
& \emph{0.002} 
& \emph{0.002} 
& 
& 0.018 
& 0.047 
& 0.004 
& 0.004 
\\ 
H & & 0.010 
& 0.008 
& 0.009 
& 0.022 
& 
& 0.014 
& 0.073 
& \emph{0.003} 
& 0.015 
\\ 
P & & 0.024 
& 0.005 
& 0.020 
& 0.005 
& 
& 0.021 
& 0.106 
& 0.022 
& \emph{0.000} 
\\ 
F & & 0.010 
& 0.006 
& 0.005 
& 0.006 
& 
& 0.320 
& 0.048 
& 0.009 
& 0.012 
\\ 
E & & 0.032 
& 0.011 
& 0.012 
& 0.004 
& 
& \emph{0.012} 
& \emph{0.022} 
& 0.010 
& 0.047 
\\ 
\hline 
\multirow{11}{*}{} \\ 

& & \multicolumn{4}{c}{Type C} & & \multicolumn{4}{c}{Type D} \\ 
\cline{3-6} \cline{8-11}
\(T^*\) & & \multicolumn{2}{c}{0.67} & \multicolumn{2}{c}{2.00} & & 
\multicolumn{2}{c}{0.67} & \multicolumn{2}{c}{2.00} \\ 
\(\eta\) & & 0.15 & 0.45 & 0.15 & 0.45 & & 
0.15 & 0.45 & 0.15 & 0.45 \\ 
\hline 
R & & 0.026 
& \emph{0.019} 
& 0.007 
& 0.008 
& 
& 0.215 
& 0.176 
& 0.024 
& \emph{0.010} 
\\ 
H & & 0.024 
& 0.045 
& \emph{0.000} 
& 0.025 
& 
& 0.211 
& \emph{0.161} 
& 0.018 
& 0.034 
\\ 
P & & \emph{0.018} 
& 0.098 
& 0.013 
& 0.016 
& 
& 0.255 
& 0.287 
& 0.019 
& 0.052 
\\ 
F & & 0.469 
& 0.054 
& 0.014 
& \emph{0.006} 
& 
& \emph{0.161} 
& 0.456 
& \emph{0.008} 
& 0.031 
\\ 
E & & 0.152 
& 0.042 
& 0.005 
& 0.062 
& 
& 0.387 
& 0.463 
& 0.040 
& 0.076 
\\ 
\hline 
\multirow{11}{*}{} \\ 

& & \multicolumn{4}{c}{Type E} & & \multicolumn{4}{c}{Type F} \\ 
\cline{3-6} \cline{8-11}
\(T^*\) & & \multicolumn{2}{c}{0.67} & \multicolumn{2}{c}{2.00} & & 
\multicolumn{2}{c}{0.67} & \multicolumn{2}{c}{2.00} \\ 
\(\eta\) & & 0.15 & 0.45 & 0.15 & 0.45 & & 
0.15 & 0.45 & 0.15 & 0.45 \\ 
\hline 
R & & --\footnotemark[1] 
& --\footnotemark[1] 
& --\footnotemark[1] 
& 0.013 
& 
& --\footnotemark[1] 
& 0.176 
& --\footnotemark[1] 
& 0.036 
\\ 
H & & --\footnotemark[1] 
& --\footnotemark[1] 
& --\footnotemark[1] 
& 0.021 
& 
& --\footnotemark[1] 
& 0.073 
& --\footnotemark[1] 
& 0.036 
\\ 
P & & --\footnotemark[1] 
& --\footnotemark[1] 
& --\footnotemark[1] 
& 0.046 
& 
& --\footnotemark[1] 
& 0.117 
& --\footnotemark[1] 
& \emph{0.022} 
\\ 
F & & --\footnotemark[1] 
& --\footnotemark[1] 
& --\footnotemark[1] 
& \emph{0.007} 
& 
& --\footnotemark[1] 
& 0.239 
& --\footnotemark[1] 
& 0.045 
\\ 
E & & --\footnotemark[1] 
& --\footnotemark[1] 
& --\footnotemark[1] 
& 0.004 
& 
& --\footnotemark[1] 
& \emph{0.032} 
& --\footnotemark[1] 
& 0.034 
\\ 
\hline 
\multirow{11}{*}{} \\ 

& & \multicolumn{4}{c}{Type G} & & \multicolumn{4}{c}{Type H} \\ 
\cline{3-6} \cline{8-11}
\(T^*\) & & \multicolumn{2}{c}{0.67} & \multicolumn{2}{c}{2.00} & & 
\multicolumn{2}{c}{0.67} & \multicolumn{2}{c}{2.00} \\ 
\(\eta\) & & 0.15 & 0.45 & 0.15 & 0.45 & & 
0.15 & 0.45 & 0.15 & 0.45 \\ 
\hline 
R & & --\footnotemark[1] 
& 0.051 
& 0.003 
& 0.040 
& 
& 0.185 
& \emph{0.068} 
& 0.012 
& 0.057 
\\ 
H & & --\footnotemark[1] 
& 0.032 
& 0.011 
& 0.041 
& 
& 0.192 
& 0.170 
& \emph{0.011} 
& 0.039 
\\ 
P & & --\footnotemark[1] 
& 0.138 
& 0.020 
& \emph{0.024} 
& 
& 0.219 
& 0.161 
& 0.020 
& 0.136 
\\ 
F & & --\footnotemark[1] 
& \emph{0.025} 
& 0.021 
& 0.036 
& 
& \emph{0.023} 
& 0.181 
& 0.014 
& \emph{0.001} 
\\ 
E & & --\footnotemark[1] 
& 0.181 
& \emph{0.001} 
& 0.049 
& 
& 0.533 
& 0.531 
& 0.027 
& 0.066 
\\ 
\end{tabular}

\end{ruledtabular}
\footnotetext[1]{Simulated system is not a single-phase, uniform fluid at equilibrium.}
\end{table}

\end{document}